\documentstyle[12pt,epsfig]{article}
\newcommand{\bm}[1]{\mbox{\boldmath $#1$}}
\newcommand{\be}{\begin{equation}}
\newcommand{\ee}{\end{equation}}
\newcommand{\bea}{\begin{eqnarray}}
\newcommand{\eea}{\end{eqnarray}}

\newcommand{\bfk}{\mbox{\boldmath $k$}}

\newcommand{\pup}{p^\uparrow}
\newcommand{\pdown}{p^\downarrow}
\newcommand{\qup}{q^\uparrow}
\newcommand{\qdown}{q^\downarrow}
\newcommand{\uup}{u^\uparrow}
\newcommand{\dup}{d^\uparrow}

\newcommand{\bfp}{\mbox{\boldmath $p$}}
\newcommand{\bfP}{\mbox{\boldmath $P$}}

\newcommand{\nd}{\noindent}
\newcommand{\la}{\lambda}
\def\lsim{\mathrel{\rlap{\lower4pt\hbox{\hskip1pt$\sim$}}\raise1pt\hbox{$<$}}}
\def\gsim{\mathrel{\rlap{\lower4pt\hbox{\hskip1pt$\sim$}}\raise1pt\hbox{$>$}}}
\def\nostrocostruttino#1\over#2{\mathrel{\mathop{\kern 0pt \rlap
{\hbox{$#1$}}} \hbox{\kern-.135em $#2$}}}
\def\sumint{\nostrocostruttino \sum \over {\displaystyle\int}}
\newcommand{\NP}[1]{{\it Nucl.\ Phys.}\ {\bf #1}}

\newcommand{\PL}[1]{{\it Phys.\ Lett.}\ {\bf #1}}
\newcommand{\PR}[1]{{\it Phys.\ Rev.}\ {\bf #1}}
\newcommand{\PRL}[1]{{\it Phys.\ Rev.\ Lett.}\ {\bf #1}}

\newcommand{\EPJ}[1]{{\it Eur.\ Phys.\ J.}\ {\bf #1}}
\newcommand{\IJMP}[1]{{\it Int.\ J.\ Mod.\ Phys.}\ {\bf #1}}

\begin{document}
\begin{flushright}
IC/HEP/04-5 \\
\end{flushright}
\vskip 1.5cm
\begin{center}
{\bf Parton intrinsic motion: suppression of the Collins mechanism for
transverse single spin asymmetries in
{\mbox{\boldmath $\pup \! p \to \pi \, X$}}}\\
\vskip 1.2cm
{\sf M.~Anselmino$^1$, M.~Boglione$^1$, U.~D'Alesio$^2$, E.~Leader$^3$,
F.~Murgia$^2$}
\vskip 0.5cm
{\it $^1$Dipartimento di Fisica Teorica, Universit\`a di Torino and \\
          INFN, Sezione di Torino, Via P. Giuria 1, I-10125 Torino, Italy}\\
\vspace{0.3cm}
{\it $^2$INFN, Sezione di Cagliari and Dipartimento di Fisica,
Universit\`a di Cagliari,\\
C.P. 170, I-09042 Monserrato (CA), Italy}\\
\vspace{0.3cm}
{\it $^3$Imperial College, Prince Consort Road, London, SW7 2BW, England}\\
\end{center}

\vspace{1.5cm}

\begin{abstract}
We consider a general formalism to compute inclusive polarised and
unpolarised cross sections within pQCD and the factorisation scheme, taking
into account parton intrinsic motion in distribution and fragmentation
functions, as well as in the elementary dynamics. Surprisingly, the intrinsic
partonic motion, with all the correct azimuthal angular dependences, produces
a strong suppression of the transverse single spin asymmetry arising from the
Collins mechanism. As a consequence, and in contradiction with
earlier claims, the Collins mechanism is unable to explain the large
asymmetries found in $\pup \, p \to \pi \, X$ at moderate to large Feynman
$x_F$. The Sivers effect is not suppressed.
\end{abstract}

\vspace{0.6cm}

\newpage
\pagestyle{plain}
\setcounter{page}{1}
\nd
{\bf 1. Introduction and general formalism}
\vskip 6pt

The inclusive production of large $p_T$ particles in the high energy collision
of two nucleons has been for a long time a crucial testing ground for
perturbative QCD; in such kinematical regions the partonic degrees of
freedom dominate the hadronic processes, which can be described in terms
of quark and gluon dynamics, coupled to non perturbative information --
parton distribution (pdf) and fragmentation (ff) functions --
gathered from other processes and evolved to the proper scale via
QCD evolution equations.

In the simplest case this translates into the well known expression:
\bea
\frac{E_C \, d\sigma^{AB \to CX}}{d^3\bfp_C} &=&
\sum_{a,b,c,d} \int dx_a \, dx_b \, dz
\, f_{a/A}(x_a, Q^2) \> f_{b/B}(x_b, Q^2)
\label{dsunp} \\
&\times& \frac{\hat s}{\pi z^2} \>
\frac{d\hat\sigma^{ab \to cd}}{d\hat t}(\hat s, \hat t, \hat u, x_a, x_b) \>
\delta(\hat s + \hat t + \hat u) \> D_{C/c}(z, Q^2) \nonumber \\
&=& \sum_{a,b,c,d} \int dx_a \, dx_b
\, f_{a/A}(x_a, Q^2) \> f_{b/B}(x_b, Q^2) \label{dsunp2} \\
&\times& \frac{1}{\pi z} \>
\frac{d\hat\sigma^{ab \to cd}}{d\hat t}(\hat s, \hat t, \hat u, x_a, x_b) \>
D_{C/c}(z, Q^2) \nonumber \>,
\eea
which combines all possible elementary QCD interactions $ab \to
cd$, with distribution, $f(x, Q^2)$, and fragmentation, $D(z,
Q^2)$, functions: all partonic intrinsic motions have been integrated 
over and the hadrons are considered as composed of collinear massless quarks
and gluons, each carrying a fraction $x$ of the parent momentum;
similarly for the final quark fragmentation into a collinear
hadron with fraction $z$ of the quark momentum. The
energy-momentum conservation of the elementary interactions, $\hat
s + \hat t + \hat u = 0$, allows to relate $x_a, x_b$ and $z$,
namely, in this collinear picture,~ $x_a \, x_b \, z \, s = -x_a
\, t - x_b \, u$, where $\hat s, \hat t, \hat u$ ($s,t,u$) are the
Mandelstam variables for the partonic (hadronic) process.

Eq. (\ref{dsunp}) -- taking into account higher order contributions to the
elementary interactions -- describes successfully the highest energy
cross section data, including the most recent ones from RHIC \cite{rhicunp}.
However, already starting from the pioneering work of Feynman, Field and Fox
\cite{fff}, several papers have shown that intrinsic transverse momenta
$\bfk_\perp$'s have to be explicitly introduced into Eq. (\ref{dsunp}) in
order to be able to explain data at moderately large $p_T$, for production of
pions and photons \cite{pkt,fu}; without them the theoretical (collinear)
computations would give results in some cases much smaller (up to a factor
10 or even more) than experiment.

Taking into account intrinsic transverse momenta is not an entirely
straightforward matter. In the pure parton model, where partons are
regarded as physical particles with definite mass (usually assumed to be
negligible), the standard collinear parton density $f_{a/A}(x_a)$ is simply
generalised to $\hat f_{a/A}(x_a,\bfk_{\perp a})$, where $\bfk_{\perp a}$
is the parton momentum perpendicular to the nucleon momentum, and
\be f_{a/A}(x_a) = \int d^2\bfk_{\perp a} \, \hat
f_{a/A}(x_a,\bfk_{\perp a}) \>, \label{fxk} 
\ee 
where, to be
precise, $x_a$ is the light-cone momentum fraction of parton $a$
inside hadron $A$.
Similarly, the fragmentation function is generalised to
$\hat D_{C/c}(z,\bfk_{\perp C})$, where $\bfk_{\perp C}$ is the transverse
momentum of the observed hadron $C$ with respect to the fragmenting parton
$c$. All dynamic partonic calculations are then carried out with inclusion
of the intrinsic transverse momenta $\bfk_\perp$'s.

This natural generalisation apparently modifies Eq. (\ref{dsunp}) into:
\bea
&& \!\!\!\!\!\! 
\frac{E_C \, d\sigma^{AB \to CX}}{d^3\bfp_C} \> = \label{dsunpgen} \\
&& \!\!\!\!\!\! 
\sum_{a,b,c,d} \int dx_a \, dx_b \, dz \, d^2\bfk_{\perp a} \,
d^2\bfk_{\perp b}\, d^3\bfk_{\perp C} \, 
\delta(\bm{k}_{\perp C} \cdot \hat{\bm{p}}_c) \> \hat f_{a/A}(x_a,
\bfk_{\perp a}; Q^2) \> \hat f_{b/B}(x_b, \bfk_{\perp b}; Q^2)
\nonumber \\
&& \!\!\!\!\!\! 
\frac{{\hat s}^2}{\pi x_a x_b z^2 s}\, J(\bm{k}_{\perp C}) 
\> \frac{d\hat\sigma^{ab \to cd}}{d\hat t}(\hat s, \hat t, \hat u, x_a, x_b) 
\> \delta(\hat s + \hat t + \hat u) \> 
\hat D_{C/c}(z, \bfk_{\perp C}; Q^2) \>,
\nonumber \eea
where $\bfk_{\perp a} \,(\bfk_{\perp b})$ and $\bfk_{\perp C}$ are
respectively the transverse momenta of parton $a \, (b)$ with
respect to hadron $A \, (B)$, and of hadron $C$ with respect to
parton $c$, which in the case  of light quarks or
gluons is taken to be massless. We have formally extended our
definition of the 2-vector $\bm{k}_{\perp C}$ into a 3-vector via
the $\delta$-function $\delta(\bm{k}_{\perp C} \cdot \hat{\bm{p}}_c)$. 
Neglecting parton masses, the function $J$ is given by \cite{fu}

\begin{equation}\label{jacobian}
J(\bm{k}_{\perp C}) = 
\frac{\left( E_C + \sqrt{\bm{p}^2_C - \bm{k}^2_{\perp C}} \right)^2}
{4(\bm{p}^2_C - \bm{k}^2_{\perp C})}
\>\cdot
\end{equation}

Eq. (\ref{dsunpgen}) has been widely used in the literature,
albeit without the factor $J$, which equals 1 if we neglect the
final hadron mass and $\bm{k}^2_{\perp C}$ in Eq.~(\ref{jacobian}). 
Note that the factor $\hat s/(\pi z^2)$ in Eq. (\ref{dsunp}) follows from 
the factor $\hat s^2/(\pi \, x_a x_b \, z^2 s)$ in (\ref{dsunpgen}) since for 
collinear collisions $\hat s = x_a x_b \, s$. Although it is true that even 
with $\bfk_\perp$, $\hat s \simeq x_a x_b \, s$, the use of this approximation 
has been shown by Cahn \cite{cahn} to lead to azimuthal asymmetries which
are physically impossible. 

The $\bfk_\perp$ dependent pdf and ff are usually assumed to have simple 
factorised and Gaussian forms, like:
\be
\hat f_{q/p}(x, \bfk_\perp; Q^2) = f_{q/p}(x, Q^2)\,g(k_\perp) =
f_{q/p}(x, Q^2) \,\frac{\beta^2}{\pi}\, e^{-\beta^2 \, k_\perp^2} \>,
\label{modu}
\ee
so that
\be
\langle k_\perp^2 \rangle =1/\beta^2 \>, \quad\quad\quad
\int d^2\bfk_\perp \, \hat f_{q/p}(x, \bfk_\perp; Q^2) = f_{q/p}(x, Q^2) \>,
\label{facpdf}
\ee
where $\beta$ might depend on $x$ and the energy; it is usually assumed
to be flavour independent. A similar factorisation is adopted for the
$\bfk_\perp$ dependent fragmentation functions. The elementary cross sections
$d\hat\sigma/d\hat t$ depend, via the elementary Mandelstam variables
$\hat s, \, \hat t$ and $\hat u$, on the intrinsic motions.

The QCD factorisation theorem implicitly used in Eq. (\ref{dsunpgen}) --
with unintegrated $\bfk_\perp$ dependent distribution and fragmentation
functions -- has never been formally proven in general \cite{col}, but only
for the Drell-Yan process, for the two-particle inclusive cross section in
$e^+e^-$ annihilation \cite{css} and, recently, for SIDIS processes in
particular kinematical regions \cite{ji}.
Moreover, in QCD the parton model is a {\it leading-twist}
approximation to the theory, whereas intrinsic transverse effects are of
{\it higher-twist} and should therefore be incorporated in a consistent
higher-twist development of the theory. Unfortunately, such a treatment is
very complicated and introduces a whole set of new unknown soft functions
and quark-gluon correlations with unclear partonic interpretation.

It turns out, however, that some partonic effects of transverse momentum are
surprisingly large and can generate phenomena which would be impossible to
reproduce in the collinear treatment:
\begin{itemize}
\item
the presence of an intrinsic $\bfk_\perp$ alters the relationship between the 
light-cone momentum fraction $x$ of the parton and the Bjorken $x_{Bj}$, so 
that $x \neq x_{Bj}$. Although the shift is small and proportional to
$\bfk_\perp^2/(x \sqrt s)^2$, it can have a substantial effect in the region
of $x$ where the parton densities are varying rapidly. This is a kind of
{\it enhanced higher-twist} effect and can lead up to an order of magnitude
change in a cross section. Similarly, due to intrinsic motion, the partonic
scattering angle in the $pp$ c.m. frame might be much smaller than the 
hadronic production angle, thus enhancing the large $p_T$ inclusive production 
of particles.
\item
In the presence of transverse momentum, certain spin-dependent effects 
can be generated by soft mechanisms and can be used to understand the large 
transverse single spin asymmetries (SSA) found in many reactions like 
$A^{\uparrow} + B \to C  + X$ and the large hyperon polarisations in 
processes like $A + B \to H^{\uparrow} + X$. At leading twist 
there are 4 such soft mechanisms, often referred to as ``odd under
naive time reversal'': \\
    a) \textit{Sivers distribution function} \cite{siv}: in a transversely
polarised nucleon with momentum $\bfp$ and polarisation vector $\bfP$, the
number density of quarks with momentum $(x\bfp,\bfk_\perp)$ is allowed to
depend upon ${\bfP} \cdot (\bfp \times \bfk_\perp)$; in other words, the
Sivers distribution function represents the azimuthal dependence (around
$\bfp$) of the number density of unpolarised quarks inside a transversely
polarised proton. \\
    b) \textit{Collins fragmentation function} \cite{col}: in the fragmentation
of a transversely polarised quark with momentum $\bfp_q$ and polarisation
vector $\bfP_{\!q}$, $q \to C + X$, the number density of hadrons $C$ with
momentum $(z\bfp_q, \bfk_\perp)$ is allowed to depend on
$\bfP_q \cdot(\bfp_q \times \bfk_\perp)$; in other words the Collins
fragmentation function represents the azimuthal dependence (around $\bfp_q$)
of the number density of unpolarised hadrons resulting from the fragmentation
of a transversely polarised quark. \\
    c) \textit{Boer-Mulders distribution function} \cite{dan}: in an
unpolarised nucleon a quark with momentum $(x\bfp,\bfk_\perp)$ is
allowed to have a non-zero polarisation along $\bfp \times
\bfk_\perp$; that is, the Boer-Mulders distribution function
represents the azimuthal dependence (around $\bfp$) of the  number
density of transversely polarised quarks inside an
unpolarised proton.\\
    d) \textit{polarising fragmentation function} \cite{pff1,pff2}: in the
fragmentation of an unpolarised quark with momentum $\bfp_q$ a final 
spin 1/2 hadron $C$ with momentum $(z\bfp_q, \bfk_\perp)$ is allowed to have 
a non-zero polarisation along $\bfp_q \times \bfk_\perp$; that is, the
polarising fragmentation function represents the azimuthal dependence
(around $\bfp_q$) of the number density of transversely polarised hadrons
resulting from the fragmentation of an unpolarised quark. \\
\end{itemize}

It should be noted that in the pure parton model, where partons
are treated as physical free particles, all these effects vanish
\cite{ecol}.

In the present paper we study transverse single spin asymmetries,
in $\pup \, p \to \pi \, X$ processes, taking into account all
parton intrinsic motions in initial and final hadrons and in the
elementary dynamics. This generalises previous work in which only the one 
$\bfk_\perp $ essential to the mechanism was taken into account, either 
in the initial polarised nucleon (Sivers effect) or in the final quark
fragmentation (Collins effect), and the $\bfk_\perp$ distribution
was somewhat simplified into essentially a two-dimensional
$\delta$-function \cite{noi1,noi2,e-e}.

For the reasons explained above, we have not attempted to
construct a fully consistent next-to-leading-twist treatment. Our
strategy is to keep only the enhanced higher-twist terms and to
calculate partonic helicity amplitudes as if the partons were
particles. We believe this approach is physically meaningful since
it takes into account the most important higher-twist terms in the
cross section and the asymmetry. Three of the above spin effects,
a)--c), can contribute to pion SSA, but in this paper we wish to
explore the generation of SSA due to the existence of the Collins
fragmentation function alone, for which there is some evidence in
the polarised lepto-production data of the HERMES collaboration
\cite{herm, herm2}. The Boer-Mulders effect can also contribute to
transverse single spin asymmetries but, at least for $\pup \, p
\to \pi \, X$ processes, it would contribute mainly at negative
$x_F$ values, whereas data are in the positive $x_F$ region. The
Sivers effect is also relevant, and has been studied in a parallel
paper \cite{fu}. In fact we shall show that the consistent
treatment of all intrinsic partonic motions induces a major
suppression of the contribution to the asymmetry due to the
Collins mechanism and renders it incapable of producing, by
itself, the kind of asymmetries measured in $\pup \, p \to \pi \,
X$ reactions \cite{e704,star}.

This result modifies the conclusions of Ref. \cite{noi2, e-e},
where the Collins contribution to SSA in $\pup \, p \to \pi \, X$
processes was computed adopting a simplified kinematical
configuration: it appeared that the Collins fragmentation function
could, although with some difficulty, explain the E704 data
\cite{e704}. Note that the results on SSA obtained using the
Sivers distribution function, with a similar simplified
kinematical configuration \cite{noi1} are, instead, essentially
confirmed by the exact treatment of all intrinsic partonic motions
\cite{fu}.

In order to study spin asymmetries we have to introduce spins in
the QCD hard scattering processes. Eq. (\ref{dsunp}) holds also
for polarised processes, $(A,S_A) + (B,S_B) \to C + X$
\cite{col2}, provided one introduces in the factorisation scheme,
in addition to the distribution functions, the helicity density
matrices which describe the parton spin states. This can be done
also for Eq. (\ref{dsunpgen}) with the result:
\bea & & \frac{E_C \, d\sigma^{(A,S_A) + (B,S_B) \to C + X}}
{d^{3} \bfp_C} = \!\!\!\!\! \sum_{a,b,c,d, \{\la\}} \> \int \frac{dx_a \,
dx_b \, dz}{16 \pi^2 x_a x_b z^2  s} \;
d^2 \bfk_{\perp a} \, d^2 \bfk_{\perp b}\, d^3 \bfk_{\perp C}\,
\delta(\bm{k}_{\perp C} \cdot \hat{\bm{p}}_c) \nonumber \\
& & J(\bm{k}_{\perp C})\, \rho_{\la^{\,}_a,
\la^{\prime}_a}^{a/A,S_A} \, \hat f_{a/A,S_A}(x_a,\bfk_{\perp a})
\> \rho_{\la^{\,}_b, \la^{\prime}_b}^{b/B,S_B} \,
\hat f_{b/B,S_B}(x_b,\bfk_{\perp b}) \label{gen1} \\
& & \hat M_{\la^{\,}_c, \la^{\,}_d; \la^{\,}_a, \la^{\,}_b} \,
\hat M^*_{\la^{\prime}_c, \la^{\,}_d; \la^{\prime}_a,
\la^{\prime}_b} \> \delta(\hat s + \hat t + \hat u) \> \hat
D^{\la^{\,}_C,\la^{\,}_C}_{\la^{\,}_c,\la^{\prime}_c}(z,\bfk_{\perp
C}) \>, \nonumber
\eea
where we have used the notation $\{\lambda\}$ to imply a sum over
{\it all} helicity indices.
In Eq.~(\ref{gen1}) $\rho_{\la^{\,}_a, \la^{\prime}_a}^{a/A,S_A}$
is the helicity density matrix of parton $a$ inside the polarised
hadron $A$ whose polarisation state is generically labelled by
$S_A$ (for spin $1/2$ particles this means longitudinal or
transverse polarisation); similarly for parton $b$ inside hadron
$B$ with spin $S_B$. The $\hat M_{\la^{\,}_c, \la^{\,}_d;
\la^{\,}_a, \la^{\,}_b}$'s are the helicity amplitudes for the
elementary process $ab \to cd$, normalised so that the unpolarised
cross section, for a collinear collision, is given by
\be
\frac{d\hat\sigma^{ab \to cd}}{d\hat t} = \frac{1}{16\pi\hat s^2}\frac{1}{4}
\sum_{\la^{\,}_a, \la^{\,}_b, \la^{\,}_c, \la^{\,}_d}
|\hat M_{\la^{\,}_c, \la^{\,}_d; \la^{\,}_a, \la^{\,}_b}|^2\,.
\label{norm}
\ee
$\hat D^{\la^{\,}_C,\la^{\prime}_C}_{\la^{\,}_c,\la^{\prime}_c}(z,
\bfk_{\perp C})$ is the product of {\it fragmentation amplitudes} for the
$c \to C + X$ process
\be
\hat D^{\la^{\,}_C,\la^{\prime}_C}_{\la^{\,}_c,\la^{\prime}_c}
= \> \sumint_{X, \la_{X}} {\hat{\cal D}}_{\la^{\,}_{X},\la^{\,}_C;
\la^{\,}_c} \, {\hat{\cal D}}^*_{\la^{\,}_{X},\la^{\prime}_C; \la^{\prime}_c}
\, , \label{framp}
\ee
where the $\sumint_{X, \la_{X}}$ stands for a spin sum and phase
space integration over all undetected particles, considered as a
system $X$. The usual unpolarised fragmentation function
$D_{C/c}(z)$, {\it i.e.} the number density
 of hadrons $C$ resulting from the fragmentation of an unpolarised
parton $c$ and carrying a light-cone momentum fraction $z$, is given by
\be
D_{C/c}(z) = {1\over 2} \sum_{\la^{\,}_c,\la^{\,}_C} \int d^2\bfk_{\perp C}
\, \hat D^{\la^{\,}_C,\la^{\,}_C}_{\la^{\,}_c,\la^{\,}_c}(z, \bfk_{\perp C})
\,. \label{fr}
\ee

Eq. (\ref{gen1}) can be formally simplified, showing its physical meaning,
by noticing that:
\be
\sum_{\la^{\,}_a, \la^{\,}_b, \la^{\prime}_a, \la^{\prime}_b, \la^{\,}_d}
\rho_{\la^{\,}_a, \la^{\prime}_a}^{A,S_A} \,
\rho_{\la^{\,}_b, \la^{\prime}_b}^{B,S_B} \,
\hat M_{\la^{\,}_c, \la^{\,}_d; \la^{\,}_a, \la^{\,}_b} \,
\hat M^*_{\la^{\prime}_c, \la^{\,}_d; \la^{\prime}_a, \la^{\prime}_b}
= \rho^{\prime}_{\la^{\,}_c, \la^{\prime}_c}(c) =
\rho_{\la^{\,}_c, \la^{\prime}_c} \, {\rm Tr}\rho^{\prime}(c) \>,
\label{rhoc}
\ee
where $\rho_{\la^{\,}_c, \la^{\prime}_c}$ is the normalised helicity density
matrix of parton $c$ produced in the $ab \to cd$ process, with initially
polarised partons $a$ and $b$; the normalisation factor Tr$\rho^{\prime}(c)$
is related to the polarised cross section for a collinear collision:
\be
{\rm Tr}\rho^{\prime}(c) = (32 \pi ^2 \hat s^2) \>
\frac{d^2\hat\sigma^{(a,s_a) + (b,s_b) \to c+d}}{d\hat t \, d \hat \phi}
\> ,
\label{normrho}
\ee
where $\hat \phi$ is the azimuthal angle of parton $c$ in the partonic
center of mass frame.
Moreover,
\be
\sum_{\la^{\,}_c\, \la^{\prime}_c, \la^{\,}_C}
\rho_{\la^{\,}_c, \la^{\prime}_c} \>
\hat D^{\la^{\,}_C,\la^{\,}_C}_{\la^{\,}_c,\la^{\prime}_c}(z,\bfk_{\perp C})
= \hat D_{C/c,s_c}(z, \bfk_{\perp C}) \>, \label{polfr}
\ee
is just the fragmentation function of a polarised parton $c$, with
spin configuration $s_c$, into a hadron $C$, whose spin is not observed.

Using Eqs. (\ref{rhoc})-(\ref{polfr}), Eq. (\ref{gen1}) can be
written as:
\bea 
& & \frac{E_C \, d\sigma^{(A,S_A) + (B,S_B) \to C + X}}
{d^{3} \bfp_C} = \sum_{a,b,c,d} \> \int dx_a \, dx_b \, dz \, d^2
\bfk_{\perp a} \, d^2 \bfk_{\perp b}\, d^3 \bfk_{\perp C}\,
\delta(\bm{k}_{\perp C} \cdot \hat{\bm{p}}_c)   \nonumber \\
& & \hat f_{a/A,S_A}(x_a,\bfk_{\perp a}) \> \hat
f_{b/B,S_B}(x_b,\bfk_{\perp b}) \> \frac{2\hat{ s}^2}{x_a x_b z^2 s}\,
J(\bm{k}_{\perp C}) \nonumber \\
& &
\frac{d^2\hat\sigma^{(a,s_a) + (b,s_b) \to c+d}}{d\hat t \, d\hat
\phi} \> \delta(\hat s + \hat t + \hat u) \> \hat D_{C/c,s_c}(z,
\bfk_{\perp C}) \>, \label{gen2} \eea
which is the analogue of Eq. (\ref{dsunpgen}) in the polarised case.

Eq. (\ref{gen2}) shows clearly the factorised structure and the partonic
interpretation: inside polarised hadrons one has polarised partons with spin
configurations $s_a$ and $s_b$, which interact via pQCD processes, leading
to a final polarised parton, with spin configuration $s_c$, which fragments
into the observed final hadron. For the initial and final step -- the
determination of the parton polarisation from the hadron polarisation and
the fragmentation of the polarised parton -- one has to rely on distribution
and fragmentation functions; some of them are known from other processes or
from theoretical models and some of them, in particular when allowing for
intrinsic motions, are new and unexplored.

Although Eq. (\ref{gen2}) has a simple physical interpretation, it is more
convenient to study the scattering process with the helicity formalism of
Eq. (\ref{gen1}); when dealing with helicities and helicity density matrices
all spins have a well defined interpretation concerning their directions
\cite{elliot}, and this is crucial if we are taking into account all parton
transverse motions, so that there are several transverse spin directions.
Since the direction of motion of the parton does not coincide with that of
its parent hadron, the longitudinal and transverse direction of the parton
spin will also be different from the longitudinal and transverse direction
of the parent hadron spin.

The partonic distribution is usually regarded, at Leading Order, as the
inclusive cross section for the process $A \to a + X$; therefore the helicity
density matrix of parton $a$ inside a hadron $A$ with polarisation $S_A$ can be
written as
\bea
\rho_{\la^{\,}_a, \la^{\prime}_a}^{a/A,S_A} \>
\hat f_{a/A,S_A}(x_a,\bfk_{\perp a})
&=& \sum_{\la^{\,}_A, \la^{\prime}_A}
\rho_{\la^{\,}_A, \la^{\prime}_A}^{A,S_A}
\sumint_{X_A, \la_{X_A}} \!\!\!\!\!
{\hat{\cal F}}_{\la^{\,}_a, \la^{\,}_{X_A};
\la^{\,}_A} \, {\hat{\cal F}}^*_{\la^{\prime}_a,\la^{\,}_{X_A}; \la^{\prime}_A}
\label{distramp} \\
&=& \sum_{\la^{\,}_A, \la^{\prime}_A}
\rho_{\la^{\,}_A, \la^{\prime}_A}^{A,S_A} \>
\hat F_{\la^{\,}_A, \la^{\prime}_A}^{\la^{\,}_a,\la^{\prime}_a} \>,\label{defF}
\eea
having defined
\be
\hat{F}_{\la^{\,}_A, \la^{\prime}_A}^{\la^{\,}_a,\la^{\prime}_a} \equiv \>
\sumint_{X_A, \la_{X_A}} \!\!\!\!\!\!
{\hat{\cal F}}_{\la^{\,}_a,\la^{\,}_{X_A};\la^{\,}_A} \,
{\hat{\cal F}}^*_{\la^{\prime}_a,\la^{\,}_{X_A}; \la^{\prime}_A} \>,
\label{defFF}
\ee
and where the $\sumint_{X_A, \la_{X_A}}\!\!\!$ stands for a spin sum and phase
space integration over all undetected remnants of hadron $A$, considered
as a system $X_A$ and the $\hat{\cal F}$'s are the {\it helicity distribution 
amplitudes} for the $A \to a + X$ process.

Notice that Eq. (\ref{defF}) relates the helicity density matrix of parton
$a$ to the helicity density matrix of hadron $A$. The helicity density matrix
describes the spin orientation of a particle in its {\it helicity rest frame}
\cite{elliot}; for a spin 1/2 particle, Tr$\,(\sigma_i \rho) = P_i$ is the
$i$-component of the polarisation vector $\bfP$ in the helicity rest frame
of the particle. In this sense Eq. (\ref{defF}) relates the hadron
polarisation to the parton polarisation, which have both to be defined
and interpreted in the proper rest frames.

The distribution function of parton $a$ inside the polarised hadron $A, S_A$
is given by
\be
\hat f_{a/A,S_A}(x_a,\bfk_{\perp a}) =
\sum_{\la^{\,}_a, \la^{\,}_A, \la^{\prime}_A}
\rho^{A,S_A}_{\la^{\,}_A, \la^{\prime}_A}
\hat F_{\la^{\,}_A, \la^{\prime}_A}^{\la^{\,}_a,\la^{\,}_a}
\, \label{pdfkp}
\ee
and the usual unpolarised distribution function $f_{a/A}(x_a)$, {\it i.e.} the
number density of partons $a$ inside an unpolarised parton $A$, carrying a
light-cone momentum fraction $x_a$, is given by
\be
f_{a/A}(x_a) = {1\over 2}
\sum_{\la^{\,}_a, \la^{\,}_A} \int d^2\bfk_{\perp a} \>
\hat F_{\la^{\,}_A, \la^{\,}_A}^{\la^{\,}_a,\la^{\,}_a}
\,. \label{pdf}
\ee

Similar expressions for the fragmentation process have already been
introduced in Eqs. (\ref{framp}) and (\ref{fr}).

By using Eq. (\ref{defF}), Eq. (\ref{gen1}) can be written as
\bea & & \hskip -30pt \frac{E_C \, d\sigma^{(A,S_A) + (B,S_B) \to C + X}}
{d^{3} \bfp_C} = \!\!\! \sum_{a,b,c,d, \{\la\}} \> \int \frac{dx_a \,
dx_b \, dz}{16 \pi^2 x_a x_b z^2 s} \;
d^2 \bfk_{\perp a} \, d^2 \bfk_{\perp b}\, d^3 \bfk_{\perp C}\,
\delta(\bm{k}_{\perp C} \cdot \hat{\bm{p}}_c) \nonumber \\
& & \hskip -30pt J(\bfk_{\perp C}) \> 
\rho_{\la^{\,}_A, \la^{\prime}_A}^{A,S_A} \>
\hat F_{\la^{\,}_A, \la^{\prime}_A}^{\la^{\,}_a,\la^{\prime}_a}
\>\>  \rho_{\la^{\,}_B, \la^{\prime}_B}^{B,S_B} \> \hat
F_{\la^{\,}_B, \la^{\prime}_B}^{\la^{\,}_b,\la^{\prime}_b} \>
\hat M_{\la^{\,}_c, \la^{\,}_d; \la^{\,}_a, \la^{\,}_b} \,
\hat M^*_{\la^{\prime}_c, \la^{\,}_d; \la^{\prime}_a,
\la^{\prime}_b} \> \delta(\hat s + \hat t + \hat u) \> \hat
D^{\la^{\,}_C,\la^{\,}_C}_{\la^{\,}_c,\la^{\prime}_c}. \label{gen3}
\eea
Eq. (\ref{gen3}) contains all possible combinations of different
distribution and fragmentation amplitudes: these combinations have
partonic interpretations and are related to the $\bfk_\perp$ and
spin dependent fragmentation and distribution functions discussed above 
and, for example, in Refs. \cite{bm} and \cite{bdr}. Notice that, even
though Eq. (\ref{dsunpgen}), for the unpolarised cross section,
looks intuitively correct and convincing, Eq. (\ref{gen3}), if
Collins and Boer-Mulders effects are operative, will yield a
different result, {\it i.e.} with $\rho_{\la^{\,}_I,
\la^{\prime}_I}^{I} = (1/2)\,\delta_{\la^{\,}_I, \la^{\prime}_I}$
($I=A,B$), Eq. (\ref{gen3}) contains terms not included in Eq.
(\ref{dsunpgen}), that is the terms off-diagonal in the parton
helicities. We have checked numerically that these contributions
are negligible in the unpolarised cross section. All this will be
discussed in detail in a forthcoming paper \cite{noi}, where all
contributions to single and double spin asymmetries will be examined, 
together with the parity and $\bfk_\perp$ properties of the distribution 
and fragmentation amplitudes. Here we are only considering the process 
$\pup \, p \to \pi \, X$ and are focussing only on the contribution of 
Collins mechanism \cite{col}, that is the azimuthal dependence of the 
number of pions created in the fragmentation of a transversely polarised
quark. The unpolarised cross section will be computed according to
Eq. (\ref{dsunpgen}), taking into account the intrinsic transverse
motion of all partons (see also Ref. \cite{fu}).

\vspace{18pt}
\goodbreak
\nd {\bf 2. Single Spin Asymmetries and Collins mechanism for pion
production } \vspace{6pt} 

Let us then consider the processes $\pup(\pdown) \, p \to \pi \, X$; we study
them in the $pp$ center of mass frame, with the polarised beam moving along
the positive $Z$-axis and the pion produced in the $XZ$ plane with 
$(\bfp_\pi)_x > 0$ values. The $\uparrow (\downarrow)$ is defined as 
the $+Y(-Y)$ direction. 
We then have, with $S_A = \, \uparrow,\downarrow$, and with an unpolarised
hadron $B$ ($S_B=0$),
\be
\rho_{\la^{\,}_A,\la^{\prime}_A}^{A,{\uparrow \downarrow}} = \frac12
\left(
\begin{array}{cc}
 1 & \mp i \nonumber \cr
 \pm i &  1 \nonumber \cr
\end{array}
\right) \quad\quad\quad
\rho_{\la^{\,}_B,\la^{\prime}_B}^{B,0} = \frac12
\left(
\begin{array}{cc}
 1 & 0 \nonumber \cr
 0 & 1 \nonumber \cr
\end{array}
\right) \> \cdot \label{rhoAB}
\ee

The computation of the single spin asymmetry
\be
A_N = \frac{d\sigma^\uparrow - d\sigma^\downarrow}
           {d\sigma^\uparrow + d\sigma^\downarrow} \label{ssa}
\ee
requires evaluation and integration, for each elementary process $ab \to cd$,
of the quantity [see Eq. (\ref{gen3})]
\be
\Sigma(S_A,S_B) \equiv  \sum_{\{\la\}} \>
\rho_{\la^{\,}_A, \la^{\prime}_A}^{A,S_A} \>
\hat F_{\la^{\,}_A, \la^{\prime}_A}^{\la^{\,}_a,\la^{\prime}_a} \>\>
\rho_{\la^{\,}_B, \la^{\prime}_B}^{B,S_B} \>
\hat F_{\la^{\,}_B, \la^{\prime}_B}^{\la^{\,}_b,\la^{\prime}_b} \>\>
\hat M_{\la^{\,}_c, \la^{\,}_d; \la^{\,}_a, \la^{\,}_b} \,
\hat M^*_{\la^{\prime}_c, \la^{\,}_d; \la^{\prime}_a, \la^{\prime}_b} \>
\hat D^{\pi}_{\la^{\,}_c,\la^{\prime}_c} \>, \label{defS}
\ee
where $\hat D^{\pi}_{\la^{\,}_c,\la^{\prime}_c}$ is defined as in
Eq. (\ref{framp}), for pion production. From Eqs. (\ref{rhoAB}) and
(\ref{defS}) one has that the numerator of $A_N$ is proportional to
\be
\Sigma(\uparrow,0) - \Sigma(\downarrow,0) = \sum_{\{\la\}} \>
\frac{(-i)}{2}
\left[
\hat F_{+,-}^{\la^{\,}_a,\la^{\prime}_a} -
\hat F_{-,+}^{\la^{\,}_a,\la^{\prime}_a}
\right] \>
\hat F_{\la^{\,}_B, \la^{\,}_B}^{\la^{\,}_b,\la^{\prime}_b} \>\>
\hat M_{\la^{\,}_c, \la^{\,}_d; \la^{\,}_a, \la^{\,}_b} \,
\hat M^*_{\la^{\prime}_c, \la^{\,}_d; \la^{\prime}_a, \la^{\prime}_b} \>
\hat D^{\pi}_{\la^{\,}_c,\la^{\prime}_c} \>, \label{delS}
\ee
while the denominator contains:
\be
\Sigma(\uparrow,0) + \Sigma(\downarrow,0) = \sum_{\{\la\}} \>
\frac{1}{2}
\left[
\hat F_{+,+}^{\la^{\,}_a,\la^{\prime}_a} +
\hat F_{-,-}^{\la^{\,}_a,\la^{\prime}_a}
\right] \>
\hat F_{\la^{\,}_B, \la^{\,}_B}^{\la^{\,}_b,\la^{\prime}_b} \>\>
\hat M_{\la^{\,}_c, \la^{\,}_d; \la^{\,}_a, \la^{\,}_b} \,
\hat M^*_{\la^{\prime}_c, \la^{\,}_d; \la^{\prime}_a, \la^{\prime}_b} \>
\hat D^{\pi}_{\la^{\,}_c,\la^{\prime}_c} \>. \label{sumS}
\ee
In the equations above and in the sequel $+$ and $-$ stand for $+1/2$ and
$-1/2$ helicities, when referring to nucleons or quarks, and for $+1$ and
$-1$ helicities, when referring to gluons.

As we have said, in this paper we are focussing solely on the
Collins mechanism and we do not consider all possible
contributions to $A_N$, which will be discussed elsewhere
\cite{noi}. Therefore, we do not consider the possibility of
finding transversely polarised quarks inside the unpolarised
proton $B$ \cite{dan} or the possibility of having different total
numbers of quarks, at different $\bfk_\perp$ values, inside the
transversely polarised proton $A$ \cite{siv}. This does not imply
that these other effects which are negligible for the unpolarised 
cross section are negligible for the SSA; simply that we wish to
explore to what extent the Collins mechanism alone is able to
explain the measured transverse single spin asymmetries. As a
consequence, the $\hat F$-terms off-diagonal in
$\la^{\,}_b,\la^{\prime}_b$ (while diagonal in
$\la^{\,}_B,\la^{\prime}_B$) and the $\hat F$-terms off-diagonal
in $\la^{\,}_A,\la^{\prime}_A$ (while diagonal in
$\la^{\,}_a,\la^{\prime}_a$) will be neglected. The Collins
mechanism corresponds to the terms off-diagonal in the fragmenting
quark helicities $\la^{\,}_c,\la^{\prime}_c$. Taking all this into
account, a partial summation in Eq. (\ref{delS}) obtains
\bea
&&\Sigma(\uparrow,0) - \Sigma(\downarrow,0) = \sum_{\{\la\}} \>
\frac{(-i)}{2} \> \biggl\{ \nonumber \\
&& \left[
\hat F_{+,-}^{+,-} - \hat F_{-,+}^{+,-}
\right]_{a/A} \> \hat f_{b/B} \>
\hat M_{\la^{\,}_c, \la^{\,}_d; +, \la^{\,}_b} \,
\hat M^*_{-\la^{\,}_c, \la^{\,}_d; -, \la^{\,}_b} \>
\hat D^{\pi}_{\la^{\,}_c,-\la^{\,}_c} \> + \label{delS1} \\
&& \left[
\hat F_{+,-}^{-,+} - \hat F_{-,+}^{-,+}
\right]_{a/A} \> \hat f_{b/B} \>
\hat M_{\la^{\,}_c, \la^{\,}_d; -, \la^{\,}_b} \,
\hat M^*_{-\la^{\,}_c, \la^{\,}_d; +, \la^{\,}_b} \>
\hat D^{\pi}_{\la^{\,}_c,-\la^{\,}_c} \> \biggr\} \>, \nonumber
\eea
where we have exploited the fact that, by parity invariance,
\be
\hat F_{+,+}^{\la^{\,}_b,\la^{\,}_b} + \hat F_{-,-}^{\la^{\,}_b,\la^{\,}_b}
= \hat f_{b/B} \>,
\ee
independently of the value of $\la^{\,}_b$.

The same procedure applied to Eq. (\ref{sumS}) reveals that the denominator
of $A_N$ is just twice the unpolarised cross section, as given by
Eq. (\ref{dsunpgen}).

Eq. (\ref{delS1}) can be further simplified by exploiting the dynamical
and the parity properties of the helicity amplitudes appearing in it. This
requires some careful considerations.

\begin{itemize}
\item
Whereas the hadronic process $\pup \, p \to \pi \, X$ takes place, according 
to our choice, in the $XZ$ plane, all other elementary processes involved: 
$A(B) \to a(b) + X$, $ab \to cd$ and $c \to \pi + X$, do not; all parton
and hadron momenta, $\bfp_a, \, \bfp_b, \, \bfp_C$, have transverse components
$\bfk_{\perp a}, \, \bfk_{\perp b}, \, \bfk_{\perp C}$
and this complicates remarkably the kinematics.
For example, the elementary QCD process $ab \to cd$, whose helicity
amplitudes are well known in the $ab$ center of mass frame, is not, in
general, a planar process anymore when observed from the $pp$ center of mass
frame. Similarly, as we commented, the spin properties described by helicity
density matrices have clear physical interpretations in each particle's own
helicity rest frame, but not necessarily in the $pp$ center of mass frame.
Of course, one can always boost and rotate from one frame into another, but
this introduces phases in the helicity amplitudes, which have to
be properly accounted for.
\item
We refer all angles to the $pp$ c.m. frame, in which $\hat{\bfp}_i =
(\theta_i, \phi_i)$, ($i=a,b,c,d$). Then the distribution functions
of the polarised proton $A$ describe processes taking place in the
plane defined by $Z$ and the $\bfp_a$ direction, $(\theta_a, \phi_a)$.
Therefore \cite{elliot, noi}
\be
\hat{\cal F}_{\la^{\,}_a,\la^{\,}_{X_A}; \la^{\,}_A}(x_a, \bfk_{\perp a})
= {\cal F}_{\la^{\,}_a,\la^{\,}_{X_A}; \la^{\,}_A}(x_a, k_{\perp a}) \>
{\rm exp}[i\la^{\,}_A \phi_a]
\label{dampphi}
\ee
and
\be
\hat F_{\la^{\,}_A,\la^{\prime}_A}^{\la^{\,}_a,\la^{\prime}_a}(x_a,
\bfk_{\perp a})
= F_{\la^{\,}_A,\la^{\prime}_A}^{\la^{\,}_a,\la^{\prime}_a}(x_a, k_{\perp a})
\> {\rm exp}[i(\la^{\,}_A - \la^{\prime}_A)\phi_a] \>, \label{fft-ff}
\ee
where $k_{\perp a} = |\bfk_{\perp a}|$;
$F_{\la^{\,}_A,\la^{\prime}_A}^{\la^{\,}_a,\la^{\prime}_a}(x_a, k_{\perp a})$
has the same definition as
$\hat{F}_{\la^{\,}_A,\la^{\prime}_A}^{\la^{\,}_a,\la^{\prime}_a}
(x_a, \bfk_{\perp a})$, Eq. (\ref{defFF}), with $\hat{\cal F}$ replaced by
${\cal F}$.

The parity properties of
${\cal F}_{\la^{\,}_a, \la^{\,}_{X_A}; \la^{\,}_A}(x_a, k_{\perp a})$ are the
usual ones valid for helicity amplitudes in the $\phi_a=0$ plane \cite{elliot},
\be
{\cal F}_{-\la^{\,}_a,-\la^{\,}_{X_A}; -\la^{\,}_A} =
\eta \, (-1)^{S_A - s_a - S_{X_A}} \> (-1)^{\la^{\,}_A - \la^{\,}_a + 
\la^{\,}_{X_A}}
\> {\cal F}_{\la^{\,}_a,\la^{\,}_{X_A}; \la^{\,}_A} \>, \label{parF}
\ee
where $\eta$ is an intrinsic parity factor such that $\eta^2 = 1$.
These imply:
\be
{F}_{-\la^{\,}_{A},-\la^{\prime}_A}^{ -\la^{\,}_a,-\la^{\prime}_{a}}
= (-1)^{2(S_A -s_a)} \>
(-1)^{(\la^{\,}_A -\la^{\,}_a) + (\la^{\prime}_A -\la^{\prime}_a)} \>
{F}_{\la^{\,}_{A},\la^{\prime}_A}^{ \la^{\,}_a,\la^{\prime}_{a}}
\,.
\label{parFF}
\ee

\item
Let us consider now the elementary partonic amplitudes. As already
remarked, the hard partonic interactions, $a(\bfp_a) + b(\bfp_b) \to
c(\bfp_c) + d(\bfp_d)$, take place out of the $XZ$ plane, which we have
chosen as the plane of the overall $\pup \, p \to \pi \, X$
process. One could compute the helicity amplitudes for these
generic processes among massless particles using techniques well
known in the literature, like those explained in Chapter 10 of
Ref. \cite{elliot}. On the other hand, the explicit expressions
and the parity properties of the helicity amplitudes $\hat M^0$,
which apply when the elementary scatterings occur in the $ab$ c.m.
frame, in the $XZ$ plane,  are well known. Therefore, rather than
computing directly the generic helicity amplitudes $\hat M$, we
prefer to relate them to the known amplitudes $\hat M^0$.

To reach the simple configuration of the $\hat M^0$ amplitudes,
starting from the generic configuration $\bfp_a, \, \bfp_b$, we
have to perform a boost in the direction determined by $(\bfp_a +
\bfp_b)$ so that the boosted three-vector $(\bfp^{\prime}_a + 
\bfp^{\prime}_b)$ is equal to zero. This will provide us with a
c.m.-like reference frame $S^\prime$ where the partons $a$ and $b$
collide head-on. Here the parton $a$ and the parton $c$,
resulting from the hard interaction between $a$ and $b$, will have
directions identified by $(\theta^\prime_a, \phi^\prime_a)$ and
$(\theta^\prime_c, \phi^\prime_c)$ respectively. In general, the
parton momenta in $S^\prime$ are related to the initial ones
(before the boost) by:
\be
\bfp_i^{\prime} = \bfp_i - \frac{\bm{q}}{q^0 + \sqrt{q^2}}
\left( \frac {p_i \cdot q}{\sqrt{q^2}} + p_i^0 \right) \label{SS'}
\ee
where $i = a,b,c,d$ and $q^\mu = (q^0, \bm{q}) = p_a^\mu + p_b^\mu$.

We need now to perform two subsequent rotations, one around the Z axis
by an angle $\phi^\prime_a$, and one around the Y axis, by an angle
$\theta^\prime_a$, such that the collision axis of the two colliding 
initial partons turns out to be aligned with the Z axis. We call this frame 
$S^{\prime\prime}$.

Under these boost and rotations the helicity states and consequently the
scattering amplitudes acquire phases, $\xi _{a,b,c,d}$ and
$\tilde \xi _{a,b,c,d}$:
\be
\hat M_{\la^{\,}_c, \la^{\,}_d; \la^{\,}_a, \la^{\,}_b}
=
\hat M^{S^{\prime\prime}} _{\la^{\,}_c, \la^{\,}_d; \la^{\,}_a, \la^{\,}_b}
\, e^{-i (\la^{\,}_a \xi _a + \la^{\,}_b \xi _b -
          \la^{\,}_c \xi _c - \la^{\,}_d \xi _d)}
\, e^{-i [(\la^{\,}_a - \la^{\,}_b) \tilde \xi _a -
         (\la^{\,}_c - \la^{\,}_d) \tilde \xi _c]} \>,
\ee
where $\xi _j$ and $\tilde \xi _j$ ($j = a,b,c,d$) are defined by
\cite{elliot}
\bea
\cos \xi _j &=&
\frac{\cos \theta _q \sin \theta _j - \sin \theta _q \cos \theta _j
      \cos(\phi _q - \phi _j)}{\sin \theta _{q p_j}} \label{cosxi} \\
\sin \xi _j &=&
\frac{\sin \theta _q \sin(\phi _q - \phi _j) }{\sin \theta _{q p_j}} \>;
\label{sinxi}
\eea
and
\be
\tilde \xi _j = \eta ^\prime _j + \xi ^\prime _j \> ,
\ee
where
\bea
\cos \xi ^\prime _j &=&
\frac{ \cos \theta _q \sin \theta ^\prime _j -
       \sin \theta  _q \cos \theta ^\prime _j
       \cos(\phi _q - \phi ^\prime _j)}
     { \sin \theta _{q p^\prime _j}} \label{cosxip} \\
\sin \xi ^\prime _j &=&
\frac{ - \sin \theta _q
       \sin(\phi _q - \phi ^\prime _j)}
     { \sin \theta _{q p^\prime _j}} \label{denxip} \>;
\eea
\bea
\cos \eta ^\prime _j &=&
\frac{\cos \theta ^\prime _a  - \cos \theta ^\prime _j
      \cos \theta _{p^\prime_a p^\prime _j} }
     {\sin \theta ^\prime _j \sin \theta _{p^\prime _a p^\prime _j}}
\label{cosetap} \\
\sin \eta ^\prime _j &=&
\frac{\sin \theta ^\prime _a \sin(\phi ^\prime _a - \phi ^\prime _j)}
     {\sin \theta _{p^\prime _a p^\prime _j}}
\label{sinetap} \>,
\eea

and the polar angles $(\theta^\prime_j , \phi^\prime_j )$ are
determined via Eq. (\ref{SS'}). Here 
$\theta _{p_i p_j}$ ($0 \leq \theta _{p_i p_j} \leq \pi$) is the angle 
between $\bfp_i$ and $\bfp_j$, and so on. Notice that $\eta ^\prime _a = 0$.

In the $S^{\prime\prime}$ frame the direction of the parton $c$ is
characterised by an azimuthal angle $\phi^{\prime\prime}_c$ given by
\be
\tan \phi^{\prime\prime}_c =
\frac{\sin \theta ^\prime _c \sin(\phi ^\prime _c - \phi^\prime_a) }
     {\sin \theta ^\prime _c \cos(\phi ^\prime _c - \phi^\prime_a)
      \cos \theta ^\prime _a -
      \cos \theta ^\prime _c \sin \theta ^\prime _a}\,\cdot
\label{phi''}
\ee
A final rotation around $Z$ of an angle $\phi^{\prime\prime}_c$ will then
finally bring us to the {\it canonical} configuration in which the partonic
process is a c.m. one in the $XZ$ plane. This introduces another phase.
As a result of the performed boost and rotations the elementary scattering
amplitudes computed in the hadronic c.m. system (the one where we are
studying the hadronic cross section) are related to the helicity amplitudes
computed in the partonic c.m. system (in the $XZ$ plane, $\phi_c=0$) by:
\be
\hat M_{\la^{\,}_c, \la^{\,}_d; \la^{\,}_a, \la^{\,}_b} \!
= \hat M^0
_{\la^{\,}_c, \la^{\,}_d; \la^{\,}_a, \la^{\,}_b}
\, e^{-i (\la^{\,}_a \xi _a + \la^{\,}_b \xi _b -
          \la^{\,}_c \xi _c - \la^{\,}_d \xi _d)}
\, e^{-i [(\la^{\,}_a - \la^{\,}_b) \tilde \xi _a -
         (\la^{\,}_c - \la^{\,}_d) \tilde \xi _c]}
\, e^{i(\la^{\,}_a - \la^{\,}_b)\phi^{\prime\prime}_c}
\label{M-M0}
\ee
with $\phi^{\prime\prime}_c$, $\xi _j$ and $\tilde \xi _j$ defined in
Eqs. (\ref{cosxi})--(\ref{phi''}); Eq. (\ref{SS'}) allows to fully express
the amplitudes in terms of the $pp$ c.m. variables $\bfp_i$. The parity
properties of the canonical c.m. amplitudes $\hat M^0$ are the usual ones:
\be
\hat M^0_{-\la^{\,}_c, -\la^{\,}_d; -\la^{\,}_a, -\la^{\,}_b} =
\eta_a \eta_b \eta_c \eta_d (-1)^{s_a + s_b - s_c - s_d} \>
(-1)^{(\la^{\,}_a - \la^{\,}_b) - (\la^{\,}_c - \la^{\,}_d)}
\hat M^0_{\la^{\,}_c, \la^{\,}_d; \la^{\,}_a, \la^{\,}_b} \>,
\label{parM0}
\ee
where $\eta_i$ is the intrinsic parity factor for particle $i$.

\item
Let us finally consider the fragmentation process. We take as
independent variables, in the $pp$ c.m. frame, the four-momentum
of the final hadron $p^{\mu}_C \equiv p^{\mu}_{\pi} =
(\sqrt{p^2_T+p^2_L},p_T,0,p_L)\,$ (whose three-momentum, 
according to our choice, lies
in the hadronic $XZ$ plane and where we neglect the pion mass), the
intrinsic transverse momentum $\bfk_{\perp C}
\equiv \bfk_{\perp \pi} = (k_{\perp \pi},\theta_{k_{\perp
\pi}},\phi_{k_{\perp \pi}})$ of the final pion with respect to
$\bfp_c$ ($\bfk_{\perp \pi} \cdot \bfp_c = 0$), and the
light-cone momentum fraction $z = p^+_\pi/p^+_c$.

The parity properties of the fragmentation amplitudes, Eq.
(\ref{framp}), are simple -- analogous to the ones for the
distribution amplitudes, Eqs. (\ref{parF}) and (\ref{parFF}) -- in
a frame $S^H$ in which the parton $c$ moves along the $Z^H$-axis.
This frame can be reached from the hadronic $pp$ frame by
performing two rotations: first around $Z$ by an angle $\phi_c$
and then around the new $Y$-axis by an angle $\theta_c$, which
brings the 3-momentum $\bm p_c$ of parton $c$ along the new 
$Z^H$-axis. In the frame $S^H$ the azimuthal angle $\phi_\pi^H$
identifying the direction of the final detected pion (which
coincides with the azimuthal angle of $\bm k_{\perp \pi}$ in
$S^H$) is given, in terms of our chosen $pp$ c.m. variables, by
\be \tan \phi^H_\pi = \pm \frac{p_T}{\sqrt{E^2_\pi-k^2_{\perp
\pi}}}\, \sqrt{1-\left(\frac{k_{\perp \pi} - p_L \cos \theta
_{k_{\perp \pi}}} {p_T \sin \theta _{k_{\perp \pi}}}\right)^2} \,
\tan \theta _{k_{\perp \pi}}\,, \label{phiC} \ee
where $E_\pi = \sqrt{p_T^2 + p_L^2}$ is the energy of the final
pion.

The analogue of Eqs. (\ref{dampphi}), (\ref{fft-ff}) and (\ref{parFF}),
for the fragmentation of a parton $c$ into a pion, reads
\be \hat{\cal D}_{\la^{\,}_{X}; \la^{\,}_c}(z, \bfk_{\perp \pi}) =
{\cal D}_{\la^{\,}_{X}; \la^{\,}_c}(z, k_{\perp \pi}) \> {\rm
exp}[i\la^{\,}_c \phi_\pi^H] \>, \label{fragphi} \ee
\be {\hat D}_{\la^{\,}_{c},\la^{\prime}_c}^{\pi}
={D}_{\la^{\,}_{c},\la^{\prime}_c}^{\pi} \> {\rm
exp}[i(\la^{\,}_{c} - \la^{\prime}_c)\phi_\pi^H] \>,
\label{ddt-dd} \ee
with the parity relationships
\be
{D}_{-\la^{\,}_{c},-\la^{\prime}_c}^{\pi}
= (-1)^{2s_c} \> (-1)^{\la^{\,}_c + \la^{\prime}_c} \>
{D}_{\la^{\,}_{c},\la^{\prime}_c}^{\pi} \,,
\label{parDD}
\ee
where ${D}_{\la^{\,}_{c},\la^{\prime}_c}^{\pi}$ is defined according to
Eq. (\ref{framp}), in the case in which the hadron $C$ is a spinless
particle (pion),
\be
D^{\pi}_{\la^{\,}_c,\la^{\prime}_c}(z, k_{\perp \pi})
= \> \sumint_{X, \la_{X}} {\cal D}_{\la^{\,}_{X};
\la^{\,}_c} \, {\cal D}^*_{\la^{\,}_{X}; \la^{\prime}_c}
\, . \label{frampn}
\ee

\end{itemize}

By exploiting the above angular and parity relations, Eqs. (\ref{fft-ff}),
(\ref{parFF}), (\ref{M-M0}), (\ref{parM0}), (\ref{ddt-dd}) and (\ref{parDD}),
we can now further simplify Eq. (\ref{delS1}). One obtains:
\bea
\Sigma(\uparrow,0) - \Sigma(\downarrow,0) &=&
-i \sum_{\{\la\}} \left[ \hat f_{b/B} \>
\hat M^0_{\la^{\,}_c, \la^{\,}_d; +, \la^{\,}_b} \>
\hat M^{0\,*}_{-\la^{\,}_c, \la^{\,}_d; -, \la^{\,}_b} \>
{D}_{\la^{\,}_{c},-\la^{\,}_c}^{\pi} \right] \label{delsimp} \nonumber \\
&& \hskip -1truecm \left\{ F_{+-}^{+-} \,
\cos [\phi_a +\phi_c^{\prime\prime} - \xi_a - \tilde \xi_a
+2\,\la^{\,}_c (\xi _c + \tilde \xi _c + \phi_\pi^H)] \right.\, \\
&& \hskip -1.3truecm - \left. \, F_{-+}^{+-} \, \cos [\phi_a
-\phi_c^{\prime\prime} + \xi_a + \tilde \xi_a -2\,\la^{\,}_c (\xi
_c + \tilde \xi _c + \phi_\pi^H)] \right\} \>, \nonumber \eea
where we have also used the fact that partons $a$ and $c$, carrying
transverse polarisation, are quarks or antiquarks, that is
$s_a = s_c = 1/2$.

Let us finally perform the remaining sum over helicities in Eq.
(\ref{delsimp}). The only types of elementary interactions contributing
are $q_a q_b \to q_c q_d$ (generically denoted as $qq$) and
$q g \to q g$ (generically denoted as $qg$), where $q_a = u,d,s,
\bar u, \bar d, \bar s$ and so on. The only independent helicity
amplitudes $\hat M^0$ for the $qq$ processes are:
\bea
&&\hat M^0_{+,+;+,+} = \hat M^0_{-,-;-,-} \equiv (\hat M_1^0)_{qq} \nonumber \\
&&\hat M^0_{-,+;-,+} = \hat M^0_{+,-;+,-} \equiv (\hat M_2^0)_{qq}
\label{Mqq} \\
&&\hat M^0_{-,+;+,-} = \hat M^0_{+,-;-,+} \equiv (\hat M_3^0)_{qq} \>.\nonumber
\eea
and, for the $qg$ processes,
\be
\hat M^0_{+,1;+,1}= \hat M^0_{-,-1;-,-1} \equiv (\hat M_1^0)_{qg} \quad\quad
 \hat M^0_{-,1;-,1} = \hat M^0_{+,-1;+,-1} \equiv (\hat M_2^0)_{qg} \>.
\label{Mqg}
\ee
At Leading Order all such amplitudes are real.

On summing over $\{\la\}$ Eq. (\ref{delsimp}) gives, for $qq$ processes,
\bea
\hskip -1truecm
\left[ \Sigma(\uparrow,0) - \Sigma(\downarrow,0) \right]_{qq} &=&
\Bigl\{ F_{+-}^{+-}(x_a, k_{\perp a}) \,
\cos [\phi_a + \phi_c^{\prime\prime} - \xi _a  - \tilde \xi _a
+ \xi _c + \tilde \xi _c + \phi_\pi^H] \nonumber \\
&-& \>\> F_{-+}^{+-}(x_a, k_{\perp a}) \, \cos [\phi_a -
\phi_c^{\prime\prime} + \xi _a  + \tilde \xi _a - \xi _c - \tilde
\xi _c - \phi_\pi^H] \Bigr\}
\label{delqq} \\
&& \hskip -1.2truecm \times \, \hat f_{q/B}(x_b, k_{\perp b}) \>
\left[ \hat M_1^0 \hat M_2^0 (x_a, x_b, z; \bfk_{\perp a},
\bfk_{\perp b}, \bfk_{\perp \pi}) \right]_{qq} \> \left[ -2i
D_{+-}^\pi(z, k_{\perp \pi}) \right] \nonumber \eea
and, for $qg$ processes
\bea
\hskip -1truecm
\left[ \Sigma(\uparrow,0) - \Sigma(\downarrow,0) \right]_{qg} &=&
\Bigl\{ F_{+-}^{+-}(x_a, k_{\perp a}) \,
\cos [\phi_a + \phi_c^{\prime\prime} - \xi _a  - \tilde \xi _a
+ \xi _c + \tilde \xi _c + \phi_\pi^H] \nonumber \\
&-& \>\> F_{-+}^{+-}(x_a, k_{\perp a}) \, \cos [\phi_a -
\phi_c^{\prime\prime} + \xi _a  + \tilde \xi _a - \xi _c - \tilde
\xi _c - \phi_\pi^H] \Bigr\}
\label{delqg} \\
&& \hskip -1.2truecm \times \, \hat f_{g/B}(x_b, k_{\perp b}) \>
\left[ \hat M_1^0 \hat M_2^0 (x_a, x_b, z; \bfk_{\perp a},
\bfk_{\perp b}, \bfk_{\perp \pi}) \right]_{qg} \> \left[ -2i
D_{+-}^\pi(z, k_{\perp \pi}) \right] \>. \nonumber \eea
The product of amplitudes appearing in Eqs.  (\ref{delqq}) and
(\ref{delqg}) are given by:
\bea
\hat M_1^0 \, \hat M_2^0 &=&
g_s^4 \, \frac89 \left\{ - \frac{\hat s \hat u}{\hat t^2}
+ \delta_{\alpha\beta} \, \frac13 \, \frac {\hat s}{\hat t}\right\}
\quad\quad\quad\> (q_\alpha q_\beta \to q_\alpha q_\beta) \nonumber \\
\hat M_1^0 \, \hat M_2^0 &=& g_s^4 \, \frac89 \, \delta_{\alpha\gamma}\left\{
- \frac{\hat s \hat u}{\hat t^2}
+ \delta_{\alpha\beta} \, \frac13 \frac {\hat u}{\hat t}\right\}
\quad\quad (q_\alpha \bar q_\beta \to q_\gamma \bar q_\delta)
\label{MM}\\
\hat M_1^0 \, \hat M_2^0 &=& g_s^4 \, \frac89 \left\{ \frac94 \,
\frac{\hat s \hat u}{\hat t^2} -1 \right\}
\quad\quad\quad\quad\quad\quad\> (qg \to qg) \nonumber
\eea
where $\alpha,\beta,\gamma$ and $\delta$ are flavour indices.
Notice that in the above expressions all the dependences on the angles
in the distribution and fragmentation functions are explicit
and the functions $F$, $\hat f$ and $D$ do not depend on angles any more;
the elementary amplitudes depend on angles via the Mandelstam variables 
$\hat s, \hat t$ and $\hat u$. Notice also that the $qq$ and $qg$ 
contributions have exactly the same structure, the difference being only in 
the parton $b$ distribution and in the elementary processes.

\mbox{}From Eqs. (\ref{gen3}), (\ref{defS}), (\ref{delqq}) and (\ref{delqg})
the numerator of the single spin asymmetry $A_N$, under the assumption that
only Collins effect contributes, is given by ($b$,$d$ can be either quarks or
gluons):
\bea
\frac{E_\pi \, d\sigma^{\pup \, p \to \pi \, X}} {d^{3} \bfp_\pi} &-&
\frac{E_\pi \, d\sigma^{\pdown \, p \to \pi \, X}} {d^{3} \bfp_\pi} =
\label{numan} \\
&& \!\!\! \sum_{q_a,b,q_c,d} \int \frac{dx_a \, dx_b \, dz}{16
\pi^2 x_a x_b z^2 s} \; d^2 \bfk_{\perp a} \, d^2 \bfk_{\perp b}\, 
d^3 \bfk_{\perp \pi}\, \delta(\bm{k}_{\perp \pi} \cdot \hat{\bm{p}}_c) 
\nonumber \\
& \times & J(\bm{k}_{\perp \pi}) \>\>
\delta(\hat s + \hat t + \hat u) \nonumber \\
&\times& \left\{ F_{+-}^{+-}(x_a, k_{\perp a}) \,
\cos [\phi_a + \phi_c^{\prime\prime} - \xi _a  - \tilde \xi _a
+ \xi _c + \tilde \xi _c + \phi_\pi^H] \right. \nonumber \\
&-& \left. \> F_{-+}^{+-}(x_a, k_{\perp a}) \,
\cos [\phi_a - \phi_c^{\prime\prime} + \xi _a  + \tilde \xi _a
- \xi _c - \tilde \xi _c - \phi_\pi^H] \right\} \nonumber \\
&&\hskip -3.3truecm \times \>\> \hat f_{b/B}(x_b, k_{\perp b}) \>
\left[ \hat M_1^0 \hat M_2^0 (x_a, x_b, z; \bfk_{\perp a},
\bfk_{\perp b}, \bfk_{\perp \pi}) \right]_{q_ab\to q_cd} \> \left[
-2i D_{+-}^\pi(z, k_{\perp \pi}) \right] \>. \nonumber \eea

A few comments are in order.
\begin{itemize}
\item
All angles appearing in Eq. (\ref{numan}) can be expressed in terms of the
$pp$ c.m. integration variables, via Eqs. (\ref{SS'}),
(\ref{cosxi})--(\ref{phi''}) and (\ref{phiC}).
\item
\mbox{}From Eqs. (\ref{parDD}) and (\ref{frampn}) one can see that
$D_{+-}^\pi$ is a purely imaginary quantity. The {\it Collins
fragmentation function} \cite{col,noi2,bdr,mul}
\be
-2 i D_{+-}^\pi = 2 \, {\rm Im} D_{+-}^\pi \equiv \Delta^ND_{\pi/\qup} \>,
\label{colf}
\ee
has a simple interpretation in the frame in which the quark moves
along the $Z$ direction, with spin parallel ($\qup$) or
antiparallel ($\qdown$) to the $Y$-axis, while the $q \to \pi \,
X$ process occurs in the $XZ$ plane: it gives the difference
between the number density of pions resulting from the
fragmentation of a quark $\qup$ and a quark $\qdown$. In the $pp$
c.m. frame the quark transverse spin direction is not, in general,
orthogonal to the $q \to \pi \, X$ plane and this reflects into
the $\phi_\pi^H$ dependence appearing in Eq. (\ref{numan}).

\item
The product of elementary amplitudes $\hat M_1^0 \hat M_2^0$, see
Eqs. (\ref{Mqq}) and (\ref{Mqg}), is, in a frame in which the partonic c.m.
scattering plane is $XZ$, simply related to the spin transfer cross section:
\be
\frac{1}{16 \pi \hat s^2} \left[ \hat M_1^0 \hat M_2^0 \right]_{qb}
= \frac{d\hat{\sigma}^{\qup \, b \to \qup \, b}}{d\hat t} -
  \frac{d\hat{\sigma}^{\qup \, b \to \qdown \, b}}{d\hat t} \> \cdot
\label{qupqd}
\ee
Again, the parton intrinsic motions give, in general, more complicated,
non planar configurations for the elementary scatterings, which induce
dependences on the angles $\xi_j$, $\xi^\prime _j$, $\eta^\prime _j$
and $\phi''_c$.
\item
The distribution terms $F_{+-}^{+-}(x_a, k_{\perp a})$ and
$F_{-+}^{+-}(x_a, k_{\perp a})$
are related to the distribution of transversely polarised quarks inside
a transversely polarised proton; these transverse directions can be
different for protons and quarks \cite{noi}. Without any intrinsic motion,
only the $F_{+-}^{+-}(x_a)$ distribution would be present,
coinciding with the transversity distribution $h_{1}(x_a)$ \cite{bdr}.

\item
Note that if one takes into account intrinsic motions only in the
fragmentation process, assumed to occur in the $XZ$ plane
[$\bfk_{\perp a} = \bfk_{\perp b}= 0$, $(\bfk_{\perp \pi})_y = 0$,
which implies all phases to be zero], one recovers the expression
for the numerator of $A_N$ (aside from the factor $J$) used in
Refs. \cite{noi2,e-e}.
\end{itemize}

We can now use Eqs. (\ref{numan}) and (\ref{dsunpgen}) to compute the SSA
$A_N = (d\sigma^\uparrow - d\sigma^\downarrow)/2\,d\sigma$.
\vspace{18pt}
\goodbreak
\nd {\bf 3. Attempts to fit the data: suppression of the Collins
mechanism}
\nobreak \vspace{6pt} \nobreak

As noted earlier, it was previously believed that the remarkably
large SSA found {\it e.g.} in the E704 experiment \cite{e704} could be
generated by either the Sivers \cite{noi1} or the Collins mechanisms
\cite{noi2,e-e}. However, to avoid handling the very complex
kinematics and having to deal numerically with 8-dimensional
integrals, only the one essential intrinsic $\bm{k}_\perp$,
responsible for the asymmetry, was taken into account in these
studies. We now believe that the phases involved, when the
kinematics is treated carefully, are crucial, and, as we shall see,
lead to a large suppression of the asymmetry due to the Collins
mechanism. As explained in \cite{fu} there is little or no
suppression of the asymmetry due to the Sivers mechanism.

In order to demonstrate the extent of the suppression we shall choose 
for the unmeasured soft functions in Eq.~(\ref{numan}) their known 
upper bounds. Let us first write these functions with the notations of 
Refs. \cite{pff1} and \cite{bdr} (details will be given in \cite{noi}): 
\bea
F_{+-}^{+-}(x, k_\perp) &=& h_1(x, k_\perp) = h_{1T}(x, k_\perp) + 
\frac{k_\perp^2}{2M_p^2} \, h_{1T}^{\perp}(x, k_\perp) \label{fmul1} \\  
F^{+-}_{-+}(x, k_\perp) &=& \frac{k_\perp^2}{2M_p^2} \, 
h_{1T}^{\perp}(x, k_\perp) \label{fmul2} \\ 
-2iD^\pi_{+-}(z, k_\perp) &=& \Delta^ND_{\pi/q^{\uparrow}}(z, k_\perp) = 
\frac{2k_\perp}{zM_\pi} \, H_1^{\perp q}(z, k_\perp) \>. \label{dmul}
\eea
where $M_p$ and $M_\pi$ are respectively the proton and pion mass.
The following positivity bounds hold \cite{sof,bbhm}: 
\bea
|h_1(x, k_\perp)| &\leq& \frac12 
\left[ q(x, k_\perp) + \Delta q(x, k_\perp) \right] = q_+(x, k_\perp)
\label{bh1} \\ 
\frac{k_\perp^2}{2M_p^2} \,
|h_{1T}^\perp(x, k_\perp)| &\leq&  \frac12 
\left[ q(x, k_\perp) - \Delta q(x, k_\perp) \right] = q_-(x, k_\perp)
\label{bh2} \\ 
|\Delta^ND_{\pi/q^{\uparrow}}(z, k_\perp)| &\leq& 2D_{\pi/q}(z, k_\perp) \>.
\label{bd1}
\eea

In our numerical estimates we adopt for all the unmeasured soft 
functions the above maximum possible values, and, moreover, adjust 
their signs so that the contributions from the valence flavours (up and down)
reinforce each other in the $\pi^+$ reaction, producing a
maximally large positive $A_N^{\pi^{+}}$. By isospin invariance
it then turns out that this choice also produces a maximally
large negative $A_N^{\pi^{-}}$. To be precise, we have computed the SSA, 
$A_N = (d\sigma^\uparrow - d\sigma^\downarrow)/2\,d\sigma$, via 
Eqs. (\ref{numan}) and (\ref{dsunpgen}), with the following choices:
\begin{itemize}
\item
For the transversity pdf $F_{+-}^{+-}(x, k_\perp) = h_1(x, k_\perp)$ and 
its companion $h_{1T}^\perp$ we have only considered up and down quark 
flavours, without any sea contribution. 
We have saturated Eqs. (\ref{bh1}) and (\ref{bh2}):
\be
h_1^u(x,k_\perp) = u_+(x,k_\perp) \quad\quad\quad 
h_1^d(x,k_\perp ) = -d_+(x,k_\perp) \label{bh3} 
\ee
\be
\frac{k_\perp^2}{2M_p^2} \, h_{1T}^{\perp u}(x,k_\perp) = 
-u_-(x,k_\perp) \quad\quad 
\frac{k_\perp^2}{2M_p^2} \, h_{1T}^{\perp d}(x,k_\perp) = 
+d_+(x,k_\perp) \>. \label{bh4} 
\ee
One naturally expects, for valence quarks, positive values for $h_1^u$ 
and negative ones for $h_1^d$; the relative signs between $h_1$ and 
$h_{1T}^\perp$ are chosen in order to maximise the sum of their 
contributions in Eq. (\ref{numan}). The $x$ and $k_\perp$ dependences 
in the unpolarised and polarised pdf are factorised assuming the same 
Gaussian form as in Eq. (\ref{modu}), with 
$\sqrt{\langle k_\perp^2} \rangle$ = 0.8 GeV/$c$ \cite{fu}. For the 
$x$-dependence of the unpolarised pdf we have adopted the MRST01 set 
\cite{mrst01} and for the polarised pdf either the LSS01 set \cite{lss01}
or the LSS-BBS set \cite{lssbbs}, as two examples of very different choices. 
We have used the same QCD evolution scale as in Ref. \cite{fu}. 

\item
We have chosen the $z$-dependence of the Collins function in such a way as 
to maximise the effects. Let us consider the production of $\pi^+$'s: since 
the dominant partonic contribution at large $x_F$ is $u g \to u g$, for which 
the product of elementary amplitudes $\hat M^0_1 \hat M^0_2$ is negative, see 
Eqs. (\ref{MM}), in order to get a positive $A_N$ we need a negative 
$u$-quark Collins function. That is, we satisfy the positivity bound 
(\ref{bd1}) with:
\be
\Delta^ND_{\pi^+/\uup}(z,k_\perp) = - 2 D_{\pi^+/u}(z,k_\perp) \>. \label{bd2}
\ee
We consider here also the contribution of the sub-leading channel
$d g \to d g$ (neglected in Refs. \cite{noi2, e-e}); as it enters with a 
negative $h_1^d$, in order to add all contributions, we use for the 
non-leading Collins function
\be
\Delta^ND_{\pi^+/\dup}(z,k_\perp) = + 2 D_{\pi^+/d}(z,k_\perp) \>. \label{bd3}
\ee
In this way also $A_N$ for $\pi^-$'s is maximised in size (by isospin
invariance).

For $\pi^0$'s we take, exploiting isospin symmetry,
\be
\Delta^ND_{\pi^0/\qup}  = 
\frac12 \left( \Delta^ND_{\pi^+/\uup} + \Delta^ND_{\pi^+/\dup} \right)
= \frac12 \left( - 2 D_{\pi^+/u} + 2 D_{\pi^+/d} \right) \>, \label{bd4} 
\ee
where $q = u, \bar u, d, \bar d$ and which still fulfills the bound 
(\ref{bd1}). The $z$ and $k_\perp$ dependences of the unpolarised 
fragmentation functions are also factorised, with the same Gaussian 
dependence as in Ref. \cite{fu}, which introduces a $z$-dependent 
$\langle k_\perp^2 \rangle$ value, smaller than the constant  
$\langle k_\perp^2 \rangle$ value assumed for the pdf. This value   
allows a good understanding of the unpolarised cross sections; we have 
explicitly checked that increasing the ff $\langle k_\perp^2 \rangle$ 
does not change significantly our present results (while spoiling the 
agreement with the unpolarised cross sections). The $z$-dependent 
unpolarised ff are taken either from Kretzer \cite{kretzer} or from 
KKP \cite{KKP,fu}, as typical examples of two different sets. 
\end{itemize}

With the above choices, Eqs. (\ref{fmul1})--(\ref{bd4}), we can (over)estimate 
the maximum value that, within our approach, the Collins mechanism alone  
contributes to the SSA in $\pup \, p \to \pi \, X$ processes.
The results are presented in the four plots of Fig.~1, which show
$(A_N)^{\rm Collins}_{\rm max}$ as a function of $x_F$, at $p_T = 1.5$ 
GeV/$c$ and $\sqrt s \simeq 19.4$ GeV: this is the E704 kinematical region and 
a comparison with their data \cite{e704} is shown. The only difference between 
the plots is given by different choices of the polarised distribution 
functions and/or the unpolarised fragmentation functions. Four different 
combinations are possible: two different sets of polarised pdf, 
LSS01 \cite{lss01} or LSS-BBS \cite{lssbbs}, and two different sets of 
unpolarised ff, Kretzer \cite{kretzer}, or KKP \cite{KKP}. The four 
combinations exhaust all possible features of choices available in the 
literature. The results clearly show that the Collins mechanism alone, even 
maximising all its effects, cannot explain the observed SSA values; its 
contribution, when all proper phases are taken into account, fails to explain 
the large E704 values observed for $A_N^{\pi^+}$ and $A_N^{\pi^-}$ at
large $x_F$. 

\vspace{18pt}
\goodbreak 
\nd
{\bf 4. Comments and conclusions}
\nobreak
\vspace{6pt}
\nobreak

We have developed a consistent formalism to describe, within pQCD and a 
factorisation scheme, the inclusive production of particles in hadronic
high energy collisions; all intrinsic motions of partons in hadrons and of
hadrons in fragmenting partons, are properly taken into account. Such a 
scheme has been applied, in a parallel paper \cite{fu}, to the description 
of several unpolarised cross sections and to the computation of SSA in
$\pup \, p \to \pi \, X$ processes, generated by the Sivers mechanism 
alone. In this paper we have again considered SSA in $\pup \, p \to \pi \, X$ 
processes, but focussing on the contribution of the Collins mechanism alone. 
Previous work \cite{noi1,noi2,e-e}, performed in a similar scheme with 
simplified kinematics, showed that both the Collins and the Sivers mechanisms, 
could alone explain the observed data on SSA. 

Such a conclusion has now to be modified: while properly chosen Sivers 
distribution functions could still explain the data \cite{fu}, there are 
no Collins fragmentation functions able to do that, as Fig. 1 shows. 
The failure of the Collins mechanism, when all partonic motions are included, 
can be understood from  the complicated azimuthal angle 
dependencies in Eq.~(\ref{numan}): the many phases arising in polarised 
distribution and fragmentation functions, and in polarised non planar 
elementary dynamics conspire, when integrated, to strongly suppress the final 
result. 

The situation with the Sivers contribution alone is much simpler, as 
the partons participating in the elementary dynamics and in the fragmentation
process are not polarised. As a consequence, the phase structure of the
numerator of $A_N$, in Sivers case, contains only one phase, the Sivers 
angle (see Eqs. (44) and (45) of Ref. \cite{fu}). Its integration, coupled 
with the dependence of the elementary dynamics on the same angle, does 
not significantly suppress the result. In this case, the simplified 
kinematics of Ref. \cite{noi1} contains the main physical features of the 
mechanism and gives a reasonably accurate computation of $A_N$.

Our results show, once more, the importance and subtleties of spin 
effects; all phases have to be properly considered and they often play 
crucial and unexpected roles. The analysis of this paper will be extended 
to other processes, like semi-inclusive Deep Inelastic Scattering, where many 
SSA effects have been observed \cite{herm,herm2} and are being measured. 

\vspace{18pt}
\goodbreak
\nd
{\bf Acknowledgements}
\vskip 6pt
The authors are grateful to INFN for continuous support to their collaboration.
M.B. is grateful to Or\'eal Italia for awarding her a grant from the 
scheme ``For Women in Science''. E.L. is grateful to the Royal Society 
of Edinburgh Auber Bequest for support. U.D. and F.M. acknowledge partial 
support from ``Cofinanziamento MURST-PRIN03''. We would like to thank 
F. Pijlman for useful discussions.

\newpage

\begin{figure} [h,t]
\epsfig{figure=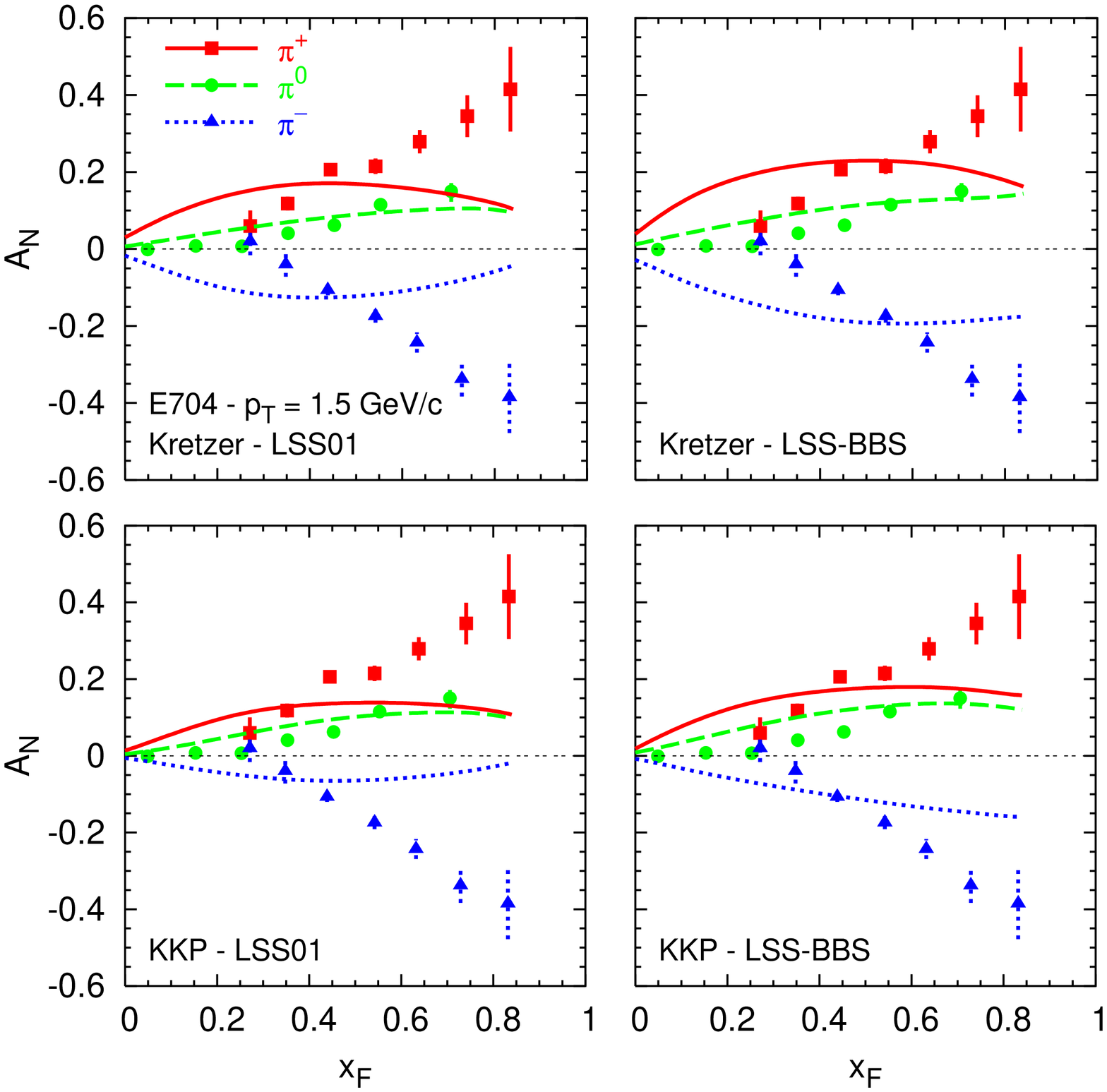,angle=-90,width=1.\textwidth}
\caption{Maximised values of $A_N$ vs.~$x_F$, 
at $\sqrt s \simeq 19.4 $ GeV and 
fixed $p_T = 1.5$ GeV/$c$, as given by the Collins mechanism 
alone; the values shown are obtained by saturating all bounds on the 
unknown soft functions and by adding constructively all different 
contributions. The four plots correspond to different choices of the 
distribution and fragmentation functions used to saturate the bounds,
as indicated in the legends. In each plot 
the upper, middle and lower sets of curves 
and data refer respectively to $\pi^+$, $\pi^0$ and $\pi^-$. Data are 
from Ref. \cite{e704}. See the text for further details.}
\label{figs}
\end{figure}

\end{document}